\documentclass[12pt,article]{elsarticle}
\def\ben{\begin{equation}}
\def\een{\end{equation}}

\def\bea{\begin{eqnarray}}
\def\eea{\end{eqnarray}}

\newcommand{\be}{\begin{eqnarray}}
\newcommand{\ee}{\end{eqnarray}}
\newcommand{\x}{x}

\newcommand{\m}{\mu}

\usepackage{xcolor}
\usepackage[colorlinks=true, pdfstartview=FitV,linkcolor=blue, citecolor=blue, urlcolor=blue]{hyperref}

\usepackage{psfrag,graphicx,xcolor}
\usepackage{latexsym,array,theorem,mathrsfs,subfigure,bm,float}
\usepackage{amsmath}
\usepackage{amsfonts}
\usepackage{amssymb}

\begin{document}

\title{Ring wormholes via duality rotations}   

\author[GG,MV]{Gary W. Gibbons}
\ead{gwg1@cam.ac.uk}

\author[MV,MV1]{Mikhail S. Volkov}
\ead{volkov@lmpt.univ-tours.fr}

\address[GG]{DAMTP, University of Cambridge, Wilberforce Road, Cambridge CB3 0WA, UK}

\address[MV]{Laboratoire de Math\'ematiques et Physique Th\'eorique,
LMPT CNRS -- UMR 7350, Universit\'e de Tours, 
Parc de Grandmont, 37200 Tours,
France
}
\address[MV1]{%
Department of General Relativity and Gravitation, Institute of Physics,\\
Kazan Federal University, Kremlevskaya street 18, 420008 Kazan, Russia
}


\begin{abstract}
We apply  duality rotations and complex transformations to the  
Schwarzschild metric to obtain wormhole geometries 
with two asymptotically flat regions connected by  a throat. 
In the simplest case these are the well-known wormholes 
supported by  phantom scalar field. Further duality rotations remove 
the scalar field to yield  less well known vacuum metrics of the oblate
Zipoy-Voorhees-Weyl class, which describe  ring wormholes. 
The ring encircles the wormhole throat and can have any radius,
whereas its tension is always negative and should be less than $-c^4/4G$. 
If the tension reaches the maximal value, the geometry becomes exactly flat,
but the topology remains non-trivial and corresponds to two copies of Minkowski space 
glued together along the disk encircled by the ring. The geodesics are 
straight lines, and those which traverse the ring get to the other universe.
The ring therefore literally produces a hole in space. 
Such wormholes could perhaps be created by negative energies concentrated  in toroidal volumes, 
for example by vacuum fluctuations.

\end{abstract}

\maketitle
Wormholes are bridges or tunnels  between
different universes or different parts of the same universe. They were first introduced 
by Einstein and Rosen  (ER) \cite{Einstein:1935tc}, who noticed that the Schwarzschild black hole 
actually has two exterior  regions connected by a bridge. The ER bridge is spacelike
and cannot be traversed by classical objects, but it has been argued   
 that it may connect quantum particles to produce quantum entanglement 
and the Einstein-Pololsky-Rosen (EPR) effect \cite{Einstein:1935rr}, hence ER=EPR \cite{Maldacena:2013xja}.
Wormholes were also considered as geometric models of elementary particles -- handles of space 
trapping inside an electric flux, say, which description  
may indeed be valid at the Planck scale  \cite{Misner:1957mt}. Wormholes can also describe initial data 
for the Einstein equations \cite {Misner:1960zz} (see \cite{Cvetic:2014vsa} for a recent review) 
whose time evolution 
corresponds to the black hole collisions of the type observed in the recent  
GW150914  event \cite{Abbott:2016blz}.

An interesting topic is traversable wormholes -- globally static bridges accessible for ordinary classical particles 
or light  \cite{Morris:1988tu} (see \cite{Visser:1995cc} for a review).  
In the simplest case such 
a wormhole is described by a static, spherically symmetric line element 
\be           \label{1}
ds^2=-Q^2(r)dt^2+dr^2+R^2(r)(d\vartheta^2+\sin^2\vartheta d\varphi^2),
\ee  
where $Q(r)$ and $R(r)$ are symmetric under $r\to -r$  and $R(r)$ 
attains a non-zero global 
minimum at $r=0$. If both $Q$ and $R/r$ approach unity as $r\to\pm\infty$ then 
the metric describes two asymptotically flat regions connected by a throat of 
radius $R(0)$.   The Einstein equations $G^\mu_\nu=T^\mu_\nu$
imply that the energy density $\rho=-T^0_0$ and the 
radial pressure $p=T^r_r$ satisfy  at $r=0$ 
\be         \label{2}
\rho+p=-2\frac{R^{\prime\prime}}{R}<0,~~~~~
p=-\frac{1}{R^2}<0.
\ee
It follows that for a static wormhole to be a solution of the Einstein equations,  
the Null Energy Condition (NEC), 
$T_{\mu\nu}v^\mu v^\nu=R_{\mu\nu} v^\mu v^\nu \geq 0$ for any null $v^\mu$, 
must be violated. 
Another  demonstration \cite{Morris:1988tu}  of the violation of the NEC uses the  
Raychaudhuri equation \cite{Hawking:1973uf} for a bundle of light rays described by
$\theta,\sigma,\omega$ : the expansion, shear and vorticity. 
In the spherically symmetric case one has  $\omega=\sigma=0$ \cite{Visser:1995cc}, 
hence 
\be        \label{4}
\frac{d\theta}{d\lambda}=-R_{\mu\nu}v^\mu v^\nu-\frac12\,\theta^2\,.
\ee
If rays pass through a wormhole throat, there is a moment of minimal cross-section area, $\theta=0$
but $d\theta/d\lambda>0$, hence $R_{\mu\nu}v^\mu v^\nu<0$ and the NEC is violated. 

If the spacetime is not spherically symmetric then the above arguments do not apply, 
but there are more subtle geometric considerations showing that the wormhole throat --
a compact two-surface of minimal area -- can exist if only the NEC is violated 
 \cite{Friedman:1993ty,Hochberg:1998ii}.
As a result,  traversable wormholes are possible if only the 
energy density becomes negative, for example due to vacuum polarization 
 \cite{Morris:1988tu}, or due to exotic matter types 
as for example phantom fields with a negative kinetic energy 
\cite{Bronnikov:1973fh,Ellis:1973yv}.
Otherwise, one can 
search for wormholes in the alternative theories of gravity, as for example 
in the Gauss-Bonnet theory \cite{Maeda:2008nz,Kanti:2011jz},
in the brainworld  models 
\cite{Bronnikov:2002rn}, 
in  theories with 
non-minimally coupled fields 
\cite{Sushkov:2011jh},
or in massive (bi)gravity \cite{Sushkov:2015fma}.

The aim of this Letter is to present a method of constructing traversable wormholes
starting from the Schwarzschild metric and applying duality rotations and complex transformations. 
As a first step this gives solutions with a non-trivial scalar field, which, after 
a  complexification, reduce to the well-known wormhole solution of Bronnikov and Ellis (BE)
\cite{Bronnikov:1973fh,Ellis:1973yv}.
A further duality rotation puts the scalar field to zero again, yielding less well known 
{\it vacuum} metrics describing ring wormholes. They  circumvent 
the above no-go arguments 
because they are not  spherically symmetric and secondly because they are not 
globally regular and exhibit a conical singularity along the ring. 
The ring encircles the wormhole throat and has a negative tension 
which should be less than $-c^4/4G$. Strikingly, 
if the tension reaches the maximal value, the geometry becomes exactly flat,
but the topology remains non-trivial and corresponds to two copies of Minkowski space 
glued together along the disk inside the ring. The geodesics are 
straight lines, and those which intersect the ring get to the other universe.
The ring therefore literally creates a hole in space. 

It is well-known that the Kerr black hole contains inside the horizon a ring singularity which 
can also be viewed as a wormhole \cite{Carter:1968rr}. 
However, its surrounding region contains closed timelike curves (CTCs). 
One should emphasize that our ring is different -- 
it is static and does not create any CTCs and moreover  is not shielded by any horizons.

\section{Model}

We consider the theory  with a scalar field, 
\be
{\cal L}=[R-2\epsilon\, (\partial \Phi)^2]\sqrt{-g}\,.
\ee
Here the parameter  takes two values, either 
$\epsilon=+1$ corresponding to the conventional scalar field,
or $\epsilon=-1$ corresponding to the phantom field. We shall denote $\Phi=\phi$
for $\epsilon=1$ and $\Phi=\psi$ for $\epsilon=-1$. 
The phantom field $\psi$ can formally be viewed as $\phi$ continued 
to imaginary values. We use units in which $G=c=1$, 
unless otherwise stated.

 Assume  the spacetime metric to be static,  
\be                    \label{static}
ds^2=-e^{2U} dt^2+e^{-2U} dl^2,
\ee
where $dl^2=\gamma_{ik}dx^i dx^k$ and 
$U$, $\gamma_{ik}$, $\Phi$ depend on the spatial coordinates $x^k$. 
Denoting by $U_k\equiv \partial_k U$, 
the Lagrangian 
becomes 
\be
{\cal L}=[\stackrel{(3)}{R}-2 \gamma^{ik}(U_i U_k+\epsilon\,\Phi_i\Phi_k)]\,\sqrt{\gamma}\,,
\ee
 where $\stackrel{(3)}{R}$ is the Ricci scalar for $\gamma_{ik}$. 
The field equations are 
\bea                    \label{eq}
\stackrel{(3)}{R}_{ik}=2(U_i U_k +\epsilon\, \Phi_i \Phi_k) \,,~~~
\Delta U=0,~~~\Delta \Phi =0.
\eea
We notice that 
for $\epsilon=1$ these equations are  invariant under 
\be                     \label{dis}
U\leftrightarrow \Phi=\phi,~~~~~~~~
\gamma_{ik}\to \gamma_{ik}\,. 
\ee
If the 3-metric $\gamma_{ik}$ is chosen to be of the Weyl form, 
\be                 \label{Weyl}
dl^2 = e^{2k} \bigl( d \rho ^2 + dz ^2 \bigr ) + \rho ^2 d \varphi ^2,
\ee
where $k$ and also $U,\Phi$ depend only on $\rho,z$, 
then the equations reduce to\footnote{One can check that this reduction is indeed consistent, which
would not be the case if the scalar field had a potential. The Weyl 
formulation is also consistent for an electrostatic vector field, so that it applies, for example,
within the electrostatic sector of  dilaton gravity.}
\bea
&&U_{\rho\rho}+\frac{1}{\rho}\,U_\rho+U_{zz}=0,~~~~~~~~~~
\Phi_{\rho\rho}+\frac{1}{\rho}\,\Phi_\rho+\Phi_{zz}=0,
  \nonumber \\
&&k_z=2\rho\,[U_\rho U_z+\epsilon\, \Phi_\rho\Phi_z],~~~~~~~
k_\rho=\rho\,[U_\rho^2-U_z^2+\epsilon\,(\Phi_\rho^2-\Phi_z^2)]. 
\eea
These equations admit a scaling symmetry mapping solutions to solutions,
\be                \label{s1}
U\to \lambda\, U,~~~~~\Phi\to \lambda\, \Phi,~~~~~k\to \lambda^2\, k,
\ee
where $\lambda$ is a constant parameter. In addition, they are invariant under
\be               \label{s2}
U\leftrightarrow \Phi,~~~~~~k\to \epsilon\, k\,.
\ee

\section{BE wormhole}
We start from the Schwarzschild metric (here $d\Omega^2=d\vartheta^2+\sin^2\vartheta d\varphi^2$)
\be
ds^2=-\left(1-\frac{2m}{r}\right)dt^2+\frac{dr^2}{1-2m/r}+r^2 d\Omega^2.
\ee
Introducing $x=r-m$ it can be put to the form \eqref{static} with 
\be                                  \label{S}
U=\frac12\ln\left(\frac{\x-m}{\x+m}\right),~~~~~~~dl^2=d\x ^2+(\x^2-m^2)\,d\Omega^2,~~~~~\Phi=0. 
\ee
Applying to this the symmetry \eqref{dis} 
 gives a new solution which 
is ultrastatic but has a non-trivial scalar field\footnote{
This is a member of the Fisher-Janis-Newman-Winicour solution family \cite{Fisher:1948yn,Janis:1968zz}.
The rest of this family can be similarly obtained from the Schwarzschild metric by applying,
instead of the discrete  symmetry  \eqref{dis}, continuous $SO(2)$ rotations in the $U,\phi$ space. 
}
\be                                  \label{SS}
U=0,~~~dl^2=d\x ^2+(\x^2-m^2)\,d\Omega^2,~~~
\phi=\frac{1}{2}\ln\left(\frac{\x-m}{\x+m}\right).
\ee
If we now continue the parameter $m$ to the imaginary region, 
$m\to i\mu$, 
then the metric remains real, while 
\be
\phi\to \frac12\ln\left(\frac{\x-i\mu}{\x+i\mu}\right)=i\psi,
\ee
where $\psi$ is real. 
This gives the solution for the phantom field, 
\be                 \label{BE}
ds^2&=&-dt^2+dx^2+(\x^2+\m^2) (d\vartheta^2+\sin^2\vartheta d\varphi^2), \nonumber  \\
\psi&=&\arctan\left(\frac{\x}{\m}\right),
\ee
which is precisely the BE wormhole 
\cite{Bronnikov:1973fh,Ellis:1973yv}.
Here $x\in(-\infty,+\infty)$ and the limits 
$x\to\pm\infty$ correspond to two asymptotically flat regions, while $x=0$ is the wormhole throat --
the 2-sphere of minimal radius $\mu$.  The geodesics can get through the throat from one 
asymptotic region to the other. 

Passing to the coordinates
\be                \label{WW}
z=x\cos\vartheta,~~\rho=\sqrt{\x^2+\m^2}\,\sin\vartheta~~\leftrightarrow~~
x\pm i\mu \cos\vartheta=\sqrt{\rho^2+(z\pm i\mu)^2}\,,
\ee
the wormhole metric assumes the Weyl form,
\be             \label{W}
ds^2=-dt^2+e^{2k}(d\rho^2+dz^2)+\rho^2d\varphi^2
\ee
with
\be
e^{2k}=\frac{\x^2+\m^2}{\x^2+\m^2\cos^2\vartheta}\,. 
\ee
Since
\be              \label{WWW}
d\rho^2+dz^2=
\left.\left.\frac{\x^2+\m^2\cos^2\vartheta}{\x^2+\m^2}
\right[d\x^2+(\x^2+\m^2) d\vartheta^2\right],
\ee
the metric \eqref{W} is indeed equivalent to the one in Eq.\eqref{BE}.

 \section{Ring wormhole}
 Let us apply to the solution \eqref{W} the symmetry 
 \eqref{s2}, $U\leftrightarrow\psi$, $k\to -k$, and then act on the result 
 with the scaling symmetry \eqref{s1} with $\lambda=\sigma$. 
This gives the axially symmetric metrics with $\Phi=0$,
\be            \label{sol}
ds^2&=&-e^{2 U}dt^2+e^{-2 U}\,dl^2,~~~~~~~~U=\sigma\arctan\left(\frac{\x}{\m}\right),  \\
dl^2&=&
\left.
\left.\left(\frac{\x^2+\m^2\cos^2\vartheta}{\x^2+\m^2}\right)^{\sigma^2+1}\right[d\x^2+(\x^2+\m^2) d\vartheta^2\right]
+(\x^2+\m^2) \sin^2\vartheta\, d\varphi^2.  \nonumber 
\ee
We note that the 3-metric $dl^2$ is invariant under $x\to -x$ while the Newtonian potential 
$U$ changes sign. 
As we shall see below, these solutions describe ring wormholes. 
We thought initially these solutions were new, but actually they were
described before although remain very little known. 

We obtained them from the BE wormhole 
by rotating away the scalar field. However, they can also be obtained from the Schwarzschild metric
without introducing any scalar fields at all. Passing to the coordinates
\be
z=x\cos\vartheta,~~\rho=\sqrt{x^2-m^2}\sin\vartheta~~\leftrightarrow~~
x\pm m\cos\vartheta=\sqrt{\rho^2+(z\pm m)^2}\,,
\ee
the Schwarzschild metric \eqref{S} assumes the Weyl form  with 
\be                                  \label{Sa}
dl^2=d\x ^2+(\x^2-m^2)\,d\Omega^2=e^{2k}(d\rho^2+d z^2)+\rho^2d\varphi^2,
\ee
where
\be
e^{2k}=\frac{x^2-m^2}{x^2-m^2 \cos^2\vartheta},
\ee
since
\be
d\rho^2+dz^2=
\left.\left.\frac{\x^2-m^2\cos^2\vartheta}{\x^2-m^2}
\right[d\x^2+(\x^2-m^2) d\vartheta^2\right].
\ee
Acting on this with the scaling symmetry \eqref{s1} with $\lambda=-\delta$ gives vacuum metrics
of the {\it prolate} Zipoy-Voorhees (ZV) class \cite{Zipoy,Voorhees:1971wh} 
\be          \label{ZV}
ds^2&=&-\left(\frac{x-m}{x+m}\right)^{-\delta} +\left(\frac{x-m}{x+m}\right)^{\delta}dl^2, \\
dl^2&=&
\left.\left.\left(\frac{\x^2-m^2\cos^2\vartheta}{\x^2-m^2}\right)^{1-\delta^2}
\right[d\x^2+(\x^2-m^2) d\vartheta^2\right]
+(\x^2-m^2) \sin^2\vartheta\, d\varphi^2\,.   \nonumber 
\ee
These metrics have been relatively well studied  
(see for example \cite{Kodama:2003ch,LukesGerakopoulos:2012pq})
since they can be used to describe deformations of the Schwarzschild metric.

If we now continue the parameters to the imaginary region, 
\be
\delta\to i\sigma,~~~~~~m\to i\mu,
\ee
then metrics \eqref{ZV} remain real and reduce to the {\it oblate} ZV solutions \cite{Zipoy,Voorhees:1971wh}, 
which are precisely the metrics \eqref{sol}. 
 Much less is known 
about these solutions. Their wormhole nature 
has been discussed \cite{Bronnikov:1997gj,Clement:1998nk} but no systematic description 
is currently available and moreover  the solutions remain largely unknown. 
We shall therefore describe  their essential properties and find new 
surprising features, notably the existence of a non-trivial flat space limit.

\section{Properties of the ring wormhole}

Similarly to the BE wormhole \eqref{BE}, 
solution \eqref{sol} has  a throat and two asymptotically flat regions. This can be most easily seen by noting 
that at the symmetry axis, where $\cos^2\vartheta=1$,
the line element $dl^2$ reduces precisely to that in \eqref{BE}, while $U$ is a bounded function.
Therefore, the 
solution indeed describes a wormhole interpolating between two asymptotically flat regions.
One has for $x\to\pm\infty$ (assuming that $\mu>0$) 
\be
e^{2U}\to e^{\pm \sigma\pi}\left(1- \frac{2\sigma\mu}{x}+\ldots\right),
\ee
hence the time coordinates in both limits differ by a factor of $e^{2\sigma\pi}$ while 
the ADM mass $M=\pm \sigma\mu$ is positive when seen from one wormhole side 
and negative from the other.

The wormhole is traversable, which can be easily seen by considering 
geodesics along the symmetry axis.  
A particle of mass ${\rm m}$ and energy $E$ follows a trajectory $x(s)$ defined by
\be
\left(\frac{dx}{ds}\right)^2+{\rm m}^2 e^{2U(x)}=E^2\,,
\ee
hence $x(s)\in (-\infty,+\infty)$  if $E^2>{\rm m}^2 e^{\sigma\pi}$. Since $U(x)$ 
grows with $x$,  
it follows that 
the wormhole attracts particles in the $x>0$ region but repels them 
in the $x<0$ region. Therefore, it acts as a ``drainhole" sucking in  matter 
from the $x>0$ region and spitting it out to the $x<0$ region.

The metric \eqref{sol} degenerates
at $x=0$, $\vartheta=\pi/2$, which corresponds to a circle of radius $\mu$. 
In its vicinity 
one can define 
$y=\m\cos\vartheta$  and then $dl^2$ in \eqref{sol} becomes 
\bea
dl^2&=&\left(\frac{x^2+y^2}{\mu^2}\right)^{\sigma^2+1}(dx^2+dy^2)+\m^2 d\varphi^2  \label{mm}  \\
&=&\mu^2\,r^{2\sigma^2+2}(dr^2+r^2d\alpha^2)+\m^2 d\varphi^2 
=\mu^2\,(dR^2+R^2d\omega^2)+\m^2 d\varphi^2,              \nonumber 
\eea
where 
$x=\mu r\cos\alpha$, $y=\mu r\sin\alpha$ while 
$R={r^{\sigma^2+2}}/{(\sigma^2 +2)}$ and $\omega=(\sigma^2+2)\,\alpha$. 
The  metric in the last line in \eqref{mm} looks flat, however, 
since $\alpha\in[0,2\pi)$, the angle $\omega$ ranges from zero to
$
\omega_{\rm max}=2\pi\,(\sigma^2+2)> 2\pi.
$
Therefore, there is a negative angle deficit 
\be
\Delta\omega=2\pi-\omega_{\rm max}=-(\sigma^2+1)\,2\pi
\ee
and hence a conical singularity at $R=0$. Since the singularity stretches in the  $\varphi$-direction,
 it sweeps  a  ring of radius $\mu$. 
Such line singularities are known to be generated by singular matter sources distributed 
along lines -- cosmic strings. Their energy per unit length    (tension) ${T}$
 is related to the angle deficit via $\Delta\omega=(8\pi G/c^4)\,{T}$
 (we restore  for a moment the correct physical dimensions),
hence the string has 
a {\it negative} tension
\be                   \label{T}
{T}=-\frac{(1+\sigma^2)\,c^4}{4G}\,.
\ee
Therefore, the wormhole solution \eqref{sol} can be viewed as sourced by a ring with  negative tension. 
The solution carries the free parameter  $\mu$, which determines the radius of the ring,
and also $\sigma$ determining the value of the ring tension. 

It is quite surprising that the self-gravitating ring can be in equilibrium and moreover 
admits an exact solution. If the string tension was positive then the string loop would be shrinking  
and the system would not be static (we are unaware 
of any exact solutions in this case, although one can construct exact 
initial data for a string loop  \cite{Frolov:1989er}).  
For the solution \eqref{sol}  the ring has a negative tension and is similar to a strut, 
hence it should rather tend to expand, which tendency is counterbalanced 
by the gravitational attraction. This must be the reason for which such an equilibrium system exists
(we do not know if the equilibrium is stable).

Even more remarkable is the following. 
Setting $\sigma=0$ in  \eqref{sol} yields 
\be            \label{sol1}
ds^2=-dt^2+
\left.\left.\frac{\x^2+\m^2\cos^2\vartheta}{\x^2+\m^2}\,\right[d\x^2+(\x^2+\m^2) d\vartheta^2\right]
+(\x^2+\m^2) \sin^2\vartheta\, d\varphi^2.
\ee
This metric is vacuum and ultrastatic hence it must be flat, so one may think this limit
is trivial. 
However, close to the symmetry axis, where $\cos^2\vartheta\approx 1$, the geometry
reduces to that for the BE wormhole \eqref{BE}.
The wormhole throat is located at $x=0$,
but, unlike in the BE case, this  is not a sphere but rather a sphere squashed to a disk with
the geometry
\be
\mu^2(\cos^2\vartheta d\vartheta^2+\sin^2\vartheta d\varphi^2)=\mu^2 (d\xi^2+\xi^2 d\varphi^2),
\ee
where $\xi=\sin\vartheta\in[0,1]$.
The disk is encircled by the ring. 

\begin{figure}[th]
\hbox to \linewidth{ \hss

	\resizebox{10cm}{9cm}{\includegraphics{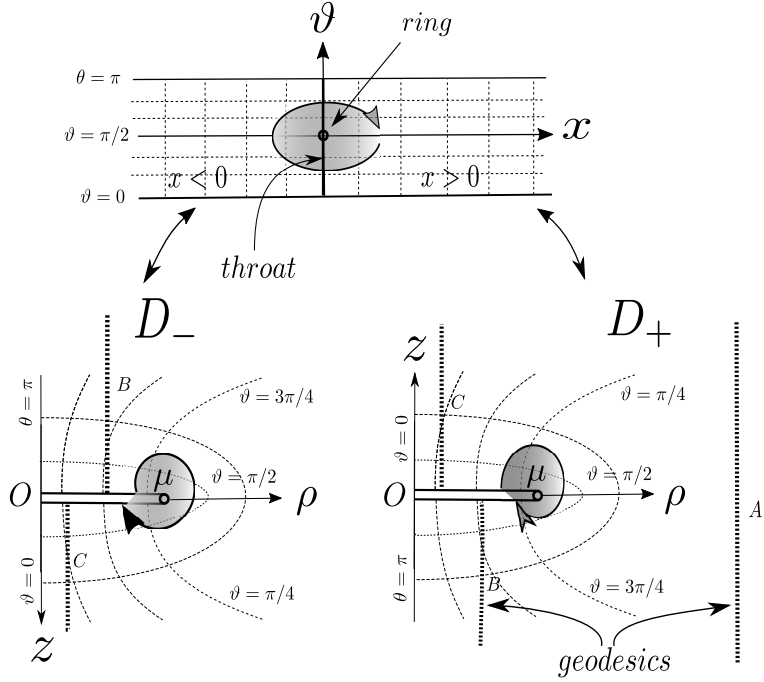}}
\hspace{1mm}
	
	
\hspace{1mm}
\hss}
\caption{Wormhole topology. The $x,\vartheta$ coordinates cover the whole of the manifold, the throat
being at $x=0$, $\vartheta\in [0,\pi]$ and the ring is at $x=0$, $\theta=\pi/2$. 
The Weyl charts $D_{+}$ and $D_{-}$ cover, respectively, the $x>0$  and $x<0$ regions.
Lines of constant $x$ are the (half)-ellipses in the Weyl coordinates, 
the ring corresponds to the branch points at $z=0,\rho=\mu$, while the throat corresponds to the branch cuts $O\mu$. 
The upper edge of the cut in the $D_{+}$ patch is identified with the lower edge of the cut in the $D_{-}$ patch 
and vice-versa.  A winding around the ring in the  $x,\vartheta$ coordinates corresponds to two windings
in Weyl coordinates. }
\label{Fig}

\end{figure}

The coordinates $x,\vartheta,\varphi$ are global, 
but to study the geodesics it is convenient to pass to the Weyl coordinates 
\eqref{WW}. The metric then becomes manifestly flat 
\be                              \label{W4}
ds^2=-dt^2+d\rho^2+dz^2+\rho^2d\varphi^2.
\ee
However, this does not mean the solution describes empty Minkowski space.
Indeed, 
the ring is still present and has the tension
\be                         \label{TT}
{T}=-\frac{c^4}{4G}\,. 
\ee
As a consequence of this the topology  is non-trivial. 
This follows from the fact that 
the Weyl coordinates  $\rho,z$  cover 
either only the $x<0$ part or only the $x>0$ part of the wormhole. Indeed, one has 
\be                        \label{ellips}
\frac{z^2}{x^2}+\frac{\rho^2}{x^2+\mu^2}=1,
\ee
hence lines of constant $x$ in the $\rho,z$ plane are ellipses 
 insensitive to the sign of $x$, 
therefore one needs two Weyl charts,
$D_\pm:\{\rho\geq 0,\, -\infty<z<\infty \}$, 
 to cover the whole of the manifold -- one for $x<0$ and one for $x>0$ 
(see Fig.\ref{Fig})  \cite{Egorov:2016rfr}. 
The coordinate transformation $(x,\vartheta)\to (\rho,z)$ degenerates at the
branch point $z=0$, $\rho=\mu$ corresponding to the position of the ring. 
Each chart therefore  has a branch cut ending at the branch point,
which may be chosen to be  
$z=0$, $\rho\in [0,\mu]$ (see Fig.\ref{Fig}). 
The upper edge of the $D_{+}$ cut is identified with the lower edge of the $D_{-}$ cut
and vice-versa. These cuts correspond to the wormhole throat, so that we see once again that 
the throat is a disk encircled by the ring.

\begin{figure}[th]
\hbox to \linewidth{ \hss

	\resizebox{7cm}{4cm}{\includegraphics{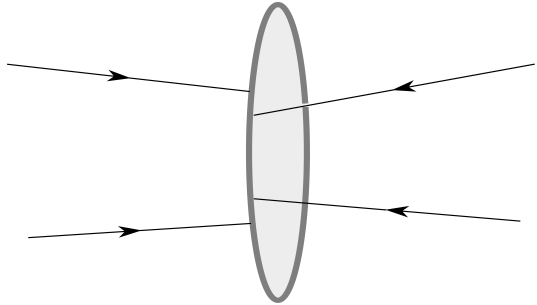}}
\hspace{1mm}
	
	
\hspace{1mm}
\hss}
\caption{Particles entering the ring are not seen coming out from the other side}
\label{Fig1}

\end{figure}

The geodesics in Weyl coordinates  are straight lines. Those which 
miss the ring always stay in the same coordinate chart (geodesic $A$ in Fig.\ref{Fig}),
while those threading the ring  (geodesics $B,C$ in Fig.\ref{Fig}) continue 
from $D_{+}$ to $D_{-}$  thus traversing the wormhole. 
 Therefore, the negative tension ring genuinely creates
a hole in space through which one can observe  another universe as well as get there%
\footnote{A similar possibility to create a hole in flat space, although without specifying the precise structure of the matter source needed, was discussed in Ref.~\cite{Krasnikov:2002dq},
following the discussion in Ref.~\cite{Visser:1995cc}.
}.
This reminds one of Alice observing the room behind the looking glass and next jumping there. 
An object falling through the ring can be seen from behind, while viewed from the side 
it is not seen coming from the other side (see Fig.\ref{Fig1}). 

It is striking  that the line source is enough to create the wormhole. Usually the negative 
energy supporting the wormhole is distributed over the 3-volume, as for the BE solution, 
or at least over a 2-surface, as for the thin-shell wormholes \cite{Garcia:2011aa}. 
We see however that it is sufficient to distribute the negative energy only along the 
one-dimensional ring, which is presumably easier to achieve than in  other cases since 
a smaller amount of the NEC violation is needed. 

It is also interesting to note that the arguments based on the 
Raychaudhuri equation do not apply 
since the wormhole either does not affect the geodesics at all 
if they miss it, or it absorbs them if they hit it 
--  its edges are sharp. 
It seems plausible that the wormhole edges could be smoothened without changing 
the global structure 
if the singular ring is replaced by a regular hoop-shaped
energy distribution  of finite  thickness and with the same tension. 
Inside the hoop the energy density is finite hence the 
geometry must be regular, while outside the energy is zero and the geometry 
should be more or less the same 
as for the original ring. 
This suggests that wormholes could be created by negative energies concentrated  in toroidal volumes, 
for example by vacuum fluctuations.

{However, the energy density needed to create a ring wormhole 
is extremely high. 
The absolute value of the negative 
tension \eqref{TT} coincides with the highest possible value for a {\it positive} tension
(force), according to 
the maximum tension principle in General Relativity conjectured in  \cite{Schiller:2006cm,Gibbons:2002iv}. 
This conjecture is supported, for example,  
by the fact that the angle deficit of a cosmic string cannot exceed $2\pi$. 
Numerically, ${T}=-3.0257\times 10^{43}$~Newtons~$\approx -3\times 10^{39}$ Tonnes.}
To create a ring of radius $R=1$ metre, say, one needs a negative energy 
equivalent to the $10^{-3}$ Solar masses, 
$2\pi RT/ c^2\approx -0.001\times M_{\odot}$. 

At the same time, one can imagine that such rings could by quantum fluctuations 
appear spontaneously from the vacuum and then disappear again. Particles crossing the 
ring during its existence would no longer be accessible from our universe after the ring disappears. 
If true, this would be a potential 
mechanism for the loss of quantum coherence.

In summary, by applying generating techniques we (re)-discovered solutions 
describing ring wormholes, computed the ring tension,  and showed that their 
geometry can be precisely flat: 
the ring creates a hole in space. 

\section*{Acknowledgements}
G.W.G. thanks the LMPT for hospitality and acknowledges the support of  ``Le Studium" -- Institute
for Advanced Studies of the Loire Valley. 
M.S.V. was partly supported by the Russian Government Program of Competitive Growth 
of the Kazan Federal University.


\end{document}